\documentclass[pra,amssymb,amsfonts,twocolumn]{revtex4}

\usepackage{amssymb}
\usepackage{euscript}
\usepackage{amsmath}
\usepackage{bm}

\def\mC {{\mathbb C}}
\def\mR {{\mathbb R}}

\def\eP{{\cal{P}}}
\def\eR{{\cal{R}}}
\def\eS{{\cal{S}}}

\def\cH{{\cal {H}}}

\def\bJ{{\bf J}}
\def\bK{{\bf K}}
\def\bP{{\bf P}}
\def\bS{{\bf S}}

\def\boa{{\bf a}}
\def\bn{{\bf n}}
\def\bp{{\bf p}}
\def\bq{{\bf q}}
\def\bs{{\bf s}}
\def\bw{{\bf w}}

\def\mSi{{{\mathbf\Sigma}}}
\def\tr{{\rm tr}}

\def\beq{\begin{equation}}
\def\eeq{\end{equation}}
\def\bea{\begin{eqnarray}}
\def\eea{\end{eqnarray}}
\def\ba{\begin{array}}
\def\ea{\end{array}}
\def\bse{\begin{subequations}}
\def\ese{\end{subequations}}
\def\tr{{\rm tr}}

\def\6{\langle }
\def\9{\rangle }
\def\half{\mbox{$1\over2$}}
\def\shalf{{\scriptstyle \frac{1}{2}}}
\def\mshalf{{\scriptstyle -\frac{1}{2}}}
\def\ra{\rightarrow}
\def\pad{\partial}
\begin{document}

\title{Two roles of relativistic spin operators}
\author {Daniel R. Terno}
\address{Department of Physics, Technion---Israel Institute of
Technology, 32000 Haifa, Israel}
\email{terno@physics.technion.ac.il.}

\begin{abstract}
Operators that are associated with several important quantities,
like angular momentum, play a double role: they are both
generators of the symmetry group and ``observables.'' The analysis
of different splittings of angular momentum into ``spin" and
``orbital" parts reveals the difference between these two roles.
We also discuss a relation of different choices of spin
observables to the violation of Bell inequalities.

\end{abstract}

\maketitle

 Spin degrees
of freedom appear in a variety of applications in quantum
information theory and foundations of quantum mechanics
\cite{per,ncb} and usually are analyzed non-relativistically. In a
relativistic domain an observable of choice is the helicity
$\bS\cdot\bp$, which is well defined for particles with sharp
momentum (for beams in accelerators typical spread to  energy
ratios are about $10^{-3}-10^{-4}$, \cite{tab}). Nevertheless,
there is also an  interest in spin operators in general,
\cite{vol,cz,fl,ve,ki}.

In this paper we consider two standard spin operators for massive
 spin-$\half$ particles, the rest frame spin and the
Dirac spin operator $\mSi$ that is associated with the spin of
moving particles as seen by a stationary observer \cite{ber}.
These two quantities can serve as  prototypes (or building blocks)
for various alternative `spin operators' that appear in the
literature
\cite{cz,fl}.

 Spin and many other operators  play a double role: they are
 symmetry generators and at the same  time are
`observables' in a sense of von Neumann measurement theory. In
most cases, both in classical and quantum physics, there is no
need to distinguish between these two roles, a notable exception
being  Koopmanian formulation of classical mechanics, where the
generators of symmetries and observables are represented by
different operators \cite{per,pt01}.

We begin from a review of necessary concepts and present a list of
properties that an operator should satisfy in order to be called
`spin'.  Then we show that even if $\mSi$ is a
discrete-degrees-of-freedom part of the generator of rotation, it
is impossible to construct one-particle Hilbert space operator
that gives the same statistics and satisfies the `spin operator'
requirements. This is similar to the analysis of van Enk and
Nienhuis of splitting  angular momentum operator of
electromagnetic field into spin and orbital parts. They show that
both are measurable quantities, but neither of them satisfies
commutation relations of angular momentum operator \cite{ve}.
Finally we discuss how a choice of the spin operator affects  a
degree of violation of the Bell-type inequalities.

For the sake of simplicity, we consider only states with a well
defined momentum. A non-zero spread in momentum has important
consequences for quantum information theory but is irrelevant for
our present subject and is described elsewhere
\cite{pst}. We set $\hbar=c=1$.

Following Wigner \cite{wig} the Hilbert space is
\beq
\cH=\mC^{2}\otimes L^2(\mR^{3},d\mu(p)),\qquad
d\mu(p)=\frac{1}{(2\pi)^3}\frac{d^3\bp}{(2p^0)},
\eeq
where $p^0=\sqrt{m^2+\bp^2}$. The generators of the Poincar\'{e}
group are represented by
\begin{subequations}
\bea
& & P^\mu = p^\mu,  \\
& & \bK=-ip^0\nabla_\bp-\frac{\bp\times\bS}{m+p^0}, \\
& & \bJ= -i\bp\times\nabla_\bp+\bS,
\eea
\end{subequations}
 where the angular momentum  is split into
orbital and spin parts, respectively.  We label basis states
$|\sigma,p\9$. Pure  state of definite momentum and arbitrary spin
will be labeled as ${\alpha\choose\beta}|p\9$.

 Lorentz transformation $\Lambda$,
$y^\mu=\Lambda^\mu_{\ \nu} x^\nu$, induces a unitary
transformation of states. In particular,
\beq
U(\Lambda)|\sigma,p\9=\sum_\xi
D_{\xi\sigma}[W(\Lambda,p)]|\xi,\Lambda p\9,\label{tran}
\eeq
where $D_{\xi\sigma}$ are the matrix elements of a unitary
operator $D$ which corresponds to a Wigner rotation
$W(\Lambda,p)$. The Wigner rotation itself  is given by
\beq
W(\Lambda,p):=L^{-1}_{\Lambda p}\Lambda L_p, \label{wigwe}
\eeq
where  $L_p$ is  a standard pure boost that takes a standard
momentum $k_R=(m,0,0,0)$ to a given momentum $p$. Explicit
formulas of $L_p$ are given, e.g., in \cite{we,bog}.

It is well known \cite{we} that for a pure rotation $\eR$ the
three-dimensional Wigner rotation matrix is the rotation itself,
\beq
W(\eR,p)=\eR,\qquad \forall p=(E(\bp),\bp). \label{pauli}
\eeq
As a result, Wigner's spin operators are nothing else but halves
of Pauli matrices (tensored with the identity of $L^2$).

A useful corollary of Eqs.~(\ref{tran}) and (\ref{wigwe}) is  a
property of the rest frame spin. If an initial state (in the rest
frame) is
\beq
|\Psi\9=\alpha|\shalf,k_R\9+\beta|{\scriptstyle -\half},k_R\9,
\eeq
with $|\alpha|^2+|\beta|^2=1$, then  a pure boost $\Lambda$  leads
to
\beq
U(\Lambda)|\Psi\9={\alpha\choose\beta}|\Lambda k_R\9.
\label{invar}
\eeq
A Pauli-Lubanski vector is an important quantity that is constructed from the
 the group generators \cite{bog},
\begin{subequations}
\bea
& & w_\varrho = \half \epsilon_{\lambda\mu\nu\varrho}P^\lambda M^{\mu\nu}, \\
& & w_0=-\bP\cdot\bJ, \qquad \bw=P^0\bJ+\bP\times\bK,
\eea
\end{subequations}
where $M^{12}=J^3, M^{01}=K^1$, etc. In particular, it helps in
splitting spin out of the angular momentum,
\beq
\bS=\frac{1}{m}\left(\bw-\frac{w^0\bP}{P^0+m}\right). \label{pl}
\eeq
For a particle with definite 4-momentum $p$, this formula is just
saying that the components of a spin operator are three spacelike
components of the Pauli-Lubanski operator at the rest frame,
\beq
S_k=(L_p^{-1}w)_k.
\eeq

We take three following properties as defining a natural
relativistic extension of the ``spin observable".
\begin{itemize}
\item The triple of operators $\bS$ reduces in the
rest frame to a non-relativistic expression $\bw_R/m$;
\item It is a three-vector
\beq
[J_j,S_k]=i\epsilon_{jkl}S_l;
\eeq
\item  It satisfies spin commutation relation
\beq
[S_j,S_k]=i\epsilon_{jkl}S_l.
\eeq
\end{itemize}
 A simple lemma (the proof can be found in
\cite{bog})  shows that this operator is
unique, under one technical assumption.

\noindent{\bf Lemma 1} The only triple of operators $\bS$ that satisfies the above
assumptions, and in addition is a linear combination of the
operators $w^\mu$ is given by Eq.~(\ref{pl}). \hfill
$\blacksquare$

To discuss Dirac spin operators we need more elements of
field-theoretical formalism. States of definite spin and momentum
 are created
from the vacuum by creation operators
$|\sigma,p\9=\hat{a}^\dag_{\sigma p}|0\9$, while antiparticles are
created by $\hat{b}^\dag_{\sigma p}$. We use the following normalisation
convention:

\beq
\6\sigma,p|\xi,q\9=(2\pi)^3
(2p^0)\delta_{\sigma\xi}\delta^{(3)}(\bp-\bq),
\eeq

Field operators are usually written with Dirac spinors.
  A Hilbert space and unitary
representation on it can be constructed from the bispinorial
representation of the Poincar\'{e} group.
 To this end we take  positive-energy solutions of Dirac equation
that form a subspace of the space of all four-component spinor
functions $\Psi=\Psi^\lambda(p)$, $\lambda=1,\dots,4$. A
Lorentz-invariant  inner product becomes positive-definite  and
the subspace of positive energy solutions becomes a Hilbert space.
The generators in this representation are
\beq
P^\mu=p^\mu,\qquad J^{\mu\nu}=L^{\mu\nu}+S^{\mu\nu}
\eeq
where
\bse
\bea
 L^{lm}=i\left(p^l\frac{\pad}{\pad p_m}-p^m\frac{\pad}{\pad
p_l}\right), & &\\
L^{0m}=ip^0\frac{\pad}{\pad p_m}\qquad
 S^{\mu\nu}=\frac{i}{4}[\gamma^\mu,\gamma^\nu]
\eea
\ese

A double infinity of plane-wave positive energy solutions of the
Dirac equation (functions $u^{(1/2)}_p$ and $u^{(-1/2)}_p$ that
are proportional to $e^{-ip\cdot x}$) is a basis of this space.
 Basis vectors of  Wigner and Dirac Hilbert spaces
are related by
\cite{bog}
\beq
u^{(1/2)}_p\Leftrightarrow|\shalf,p\9,\qquad
u^{(-1/2)}_p\Leftrightarrow|\mshalf,p\9,
\eeq

A discrete part of $S^{\mu\nu}$ ($\Sigma^1/2\equiv S^{23}$, etc)
is a Dirac spin operator. In standard or Weyl representations it
looks like
\beq
\mSi=\left(
\ba{cc}
\mbox{\boldmath $\sigma$} & 0\\
0 & \mbox{\boldmath $\sigma$}
\ea\right).
\eeq

It is possible to say that different propositions for spin
operators are different ways to split $\bJ$. However, a
momentum-dependent Foldy-Wouthuysen transformation takes
$\mbox{\boldmath $\sigma$}$ to the spinor representation of $\bS$,
\cite{ber,bog}. Hence we see that the difference is essentially
in a covariant treatment.

A field operator is constructed with the help of plane wave
solutions of Dirac equations \cite{ber,we,bog},
\beq
\hat{\phi}(x)=\sum_{\sigma=\half,-\half}\int\!d\mu(p)(e^{ix\cdot
p}v^\sigma_p\hat{b}^\dag_{\sigma p}+e^{-ix\cdot
p}u^\sigma_p\hat{a}_{\sigma p}),
\eeq
where $v^\sigma_p$ are negative energy plane wave solutions of
Dirac equation.

Using field transformation properties it is a standard exercise to
get the following expression for Dirac spin operator \cite{jauch}
\bea
& & \hat{\mSi}=::\int d^3x
\hat{\phi}^\dag(x)\frac{\mSi}{2}\hat{\phi}(x)::
\label{spid}
\eea
where :: designates a normal ordering. Wigner spin is given by
\beq
\hat{\bS}=\half\sum_{\eta,\zeta}\mbox{\boldmath
$\sigma$}_{\eta\zeta}\int\!d\mu(p) (\hat{a}^\dag_{\eta
p}\hat{a}_{\zeta p}+\hat{b}^\dag_{\eta p}
\hat{b}_{\sigma p}).
\eeq

An interpretation of $\hat{\mSi}$ and $\hat{\bS}$ as observables  is based on
 the analysis of one-particle states with
well-defined momentum and an arbitrary spin, such as
$|\Psi\9={\alpha\choose\beta}|p\9\equiv\psi|p\9$. A corresponding
Dirac spinor for this state is  $\Psi_p=\alpha u^{(1/2)}_p+\beta
u^{(-1/2)}_p$.

An expectation value of Wigner spin operator is just a
non-relativistic rest-frame expression
\beq
\bar{\bs}=\frac{\6\Psi|\hat{\bS}|\Psi\9}{\6\Psi|\Psi\9}=\psi^\dag\frac{\mbox{\boldmath
$\sigma$}}{2}\psi
\eeq
 The transformation properties of momentum
eigenstates Eq.~(\ref{invar}), Lemma 1
and the fact that Wigner spin operator commutes with the
Hamiltonian lead to the association of $\hat{\bS}$ with a
conserved quantity `rest frame spin'.

 Dirac spin $\mSi$ is
associated with the spin of a moving particle. A quantity
\beq
\bar\bs^D=\frac{\6\Psi|\hat{\mSi}|\Psi\9}{\6\Psi|\Psi\9}=
\Psi_p^\dag\frac{\mSi}{4E(\bp)} \Psi_p \label{dispin}
\eeq
is interpreted as an expectation value of the spin of a moving particle
with momentum
$\bp$ \cite{ber}. It reduces to its non-relativistic value for $\bp\ra 0$ and
particle's helicity can be calculated with either of the operators.

While from the Lemma 1 it is clear that $\mSi$ does not define
spin operators on the one-particle Hilbert space, it is
instructive to see how it fails to do so. En route we construct
$\mbox{\boldmath $\eS$}$, an one-particle Hilbert space
restriction of $\hat{\mSi}$. To this end we derive a necessary and
sufficient condition for three expectation values to be derivable
from the three operators that satisfy spin commutation relations.

Consider six $2\times 2$ spin density matrices with
 Bloch vectors $\pm\hat{\bf z}, \pm\hat{\bf  x}$ and $\pm\hat{\bf y}$.
 These density matrices are
\beq
\rho_z=\left(
\ba{cc}
1 & 0 \\ 0 & 0\ea\right), \rho_x=\left(
\ba{cc}
\half & \half \\ \half & \half\ea\right), \rho_y=\left(
\ba{cr}
\half & -\frac{i}{2} \\ \frac{i}{2} & \half\ea\right),
\eeq
etc. We are looking for three Hermitian $2\times 2$ matrices
$\eS_k$ the expectation values of which on the above states are
the prescribed numbers $\bar{s}_k(\rho_l)$,
\beq
\bar{s}_k(\rho_l)=\tr(\eS_k\rho_l),\qquad k=1,2,3;~~~
\pm l=\pm x,\pm y,\pm z.\label{def}
\eeq
We decompose these matrices in terms of Pauli matrices,
\beq
\eS_k=\sum_{n=0}^3s_{kn}\sigma_n,
\eeq
where $\sigma_0$ is an identity. It is easy to see that
\beq
\tr(\rho_l\sigma_n)=\delta_{ln},\hspace{10mm} n=1,2,3.
\eeq
Therefore, $ \bar{s}_k(\rho_l)=s_{k0}+s_{kl}$. If instead of spin
states $\rho_l$ we take their orthogonal complements $\rho_{-l}$
 we
see that all $s_{k0}=0$, so
\beq
\eS_k=\sum_l\bar{s}_{k}(\rho_l)\sigma_l.\label{reco}
\eeq
We want these operators to satisfy spin commutation relations
$[\eS_j,\eS_k]=i\epsilon_{jkl}\eS_l.$
Therefore,
\beq
\bar{s}_j(\rho_m)\bar{s}_k(\rho_n)[\sigma_m,\sigma_n]=2i\epsilon_{jkl}
\bar{s}_l(\rho_p)\sigma_p,
\eeq
holds and  the summation is understood over the repeated indices.
As a result we establish the following

\noindent{\bf Lemma 2} A necessary and sufficient condition for a triple of
probability distributions on spin-$\half$ states with the expectation values
$\bar{\bs}=(\bar{s}_1,\bar{s}_2,\bar{s}_3)$ to be derived from a triple of
 matrices that satisfy spin commutation relations 
 is
\beq
\bar{s}_j(\rho_m)\bar{s}_k(\rho_n)\epsilon_{mnp}=
\epsilon_{jkl}\bar{s}_l(\rho_p)
\eeq
where the three states $\rho_p$ are the pure states with Bloch vectors
$\hat{\bf x}$, $\hat{\bf y}$ and $\hat{\bf z}$, respectively
 \hfill $\blacksquare$

We apply this technique to a relativistic spin. Six  states $\rho$ are taken to be
spin parts of  zero momentum states.
 Consider them in a frame where they have a momentum
$\bp=p\,(n_x,n_y,n_z)$. Expectation values of $\bs^D$ in that
frame are calculated according to Eq.~(\ref{dispin}). For $\rho_z$
we get
\begin{widetext}
\beq
 \bar{\bs}^D(\rho_z)=\frac{1}{2} \frac{1}{p^2 + m(m + \sqrt{m^2 + p^2})}
  \left( n_xn_zp^2,n_y n_zp^2,m^2 + n_z^2p^2 + m{\sqrt{m^2 +
  p^2}}\right),
\eeq
\end{widetext}
and analogous expressions for other states. From Eq.~(\ref{reco})
we see that $\mbox{\boldmath $\eS$}$ is given by
\beq
\eS_k=\sum_l\bar{s}^D_{k}(\rho_l)\sigma_l
\eeq
However, a simple calculation reveals that, e.g.,
\beq
[\eS_x(\bp),\eS_y(\bp)]\neq i\eS_z(\bp),
\eeq
and the equality is recovered only in the non-relativistic limit.
It is not just a problem of a combining three operators into
3-vector. If we write $n_z=\cos\theta$ we find that the
eigenvalues of $\eS_z$ are
\beq
s_\pm=\pm\frac{1}{2}\frac{\sqrt{E^2(\bp)+m^2+\bp^2 \cos2\theta}}
{\sqrt{2}E(\bp)}. \label{eigen}
\eeq

A triple of operators $\eS_j\otimes 1_{L^2}$ is a  restriction of
$\hat{\mSi}$  that operates on the  Fock space
${\cal{F(H)}}=\oplus_n S^-(\cH^{\otimes n})$ to the one-particle
space $\cH$. In the process of restriction the essential spin
operator properties are lost, even if the resulting operators are
the legitimate observables, similarly to \cite{ve}.

 If one requires fixed outcomes $\pm\half$ they are possible to achieve
 at the
price of introducing a two-outcome positive operator-valued
measure (POVM)
\cite{per,ncb}. The expectation value $\bar{s}_k^D$ implies that
the probabilities of the outcomes $\pm\half$ are
\beq
p_k^\pm(\rho_l)=\half(1\pm 2\bar{s}^D_k(\rho_l)).
\eeq
 Using the operators $\eS_k$ we  can
construct  projectors $\eP^\pm_k$ on the one-particle Hilbert
space that correspond to projectors $\hat{P}^\pm_k$ on the Fock
space. They are
\beq
\eP^\pm_k=\half(1\pm 2\eS_k).
\eeq
By a simple inspection we find that, e. g.,
\beq
\eP^+_z\eP^-_z\neq 0, \label{bell}
\eeq
and the orthogonality is recovered only in the limit $\bp\ra 0$.
Since $\eP^\pm_k>0$ and $\eP^+_k+\eP^-_k=1$ we indeed have a
two-outcome POVM.

 From these results we learn that being a representation of
 a symmetry generator
does not necessarily imply that this operator is also an
observable with `usual' properties. We have two distinct
representations of SU(2) algebra on the spin-$\half$ Fock space,
$\hat{\bS}$ and $\hat{\mSi}$. However, only one of them preserves
defining commutation relations when restricted to one-particle
Hilbert space.

Now let us consider a relation of different spin operators to the
maximal violation of Bell inequalities
\cite{per,ncb}. Consider the Clauser-Horne
\cite{ch} version of Bell inequalities, where two pairs of
operators describe pairs of possible tests ($A_1$ and $A_2$ for
the first particle, $B_1$ and $B_2$ for the second). In each test
two possible outcomes are conventionally labeled  `+' and `-'.
Probabilities of these outcomes, e. g, for the first particle,
 are given as expectations $p_i^\pm=\tr(E_i^\pm\rho)$,
where positive operators $E_i^\pm$ form a two-outcome POVM,
$E_i^++E_i^-=1$. The four operators $A_i$, $B_i$ are defined
similarly to Eq.~(\ref{bell}). In particular, $A_i=2E^+_i-1$, and
the absence of a factor $\half$ is conventional.

It was shown by Summers and Werner
\cite{sw} that the inequalities are maximally violated only if each couple
of operators generates spin commutation relations. In particular,
the operators $A_i$ have to satisfy $A_i^2=1$ and
$A_1A_2+A_2A_1=0$, and  operators $B_i$ are similarly constrained.
Hence, defining $A_3:=-\frac{i}{2}[A_1,A_2]$ one indeed reproduces
commutation relations of Pauli matrices.

Now assume that these operators are realized as
$A_i=2\boa_i\cdot\mbox{\boldmath $\eS$}$, etc., where ${\boa}_i$
is a unit vector.  Then Eq.~(\ref{eigen}) shows that generically
there will be less than maximal violations of the inequalities.

Czachor \cite{cz} considers a different spin operator,
$\tilde{\bS}$, which is suitably normalized Pauli-Lubanski
operator $\bw$. Then $A_i=2\boa_i\cdot\tilde{\bS}$, so
\beq
A_i=2\left[\frac{m}{p^0}{\boa}_i+
\left(1-\frac{m}{p^0}\right)(\boa\cdot\bn)\bn\right]\cdot\bS
\equiv2\mbox{\boldmath $\alpha$}(\boa,\bp)\cdot\bS,
\eeq
where $\bS$ is the Wigner spin operator and $\bn=\bp/|\bp|$. The
length of the auxiliary vector $\mbox{\boldmath $\alpha$}$ is
\beq
|\mbox{\boldmath$\alpha$}|=
\frac{\sqrt{(\bp\cdot\boa)^2+m^2}}{p^0},
\eeq
so we see that generically $A_i^2=\mbox{\boldmath $\alpha$}^21<1$.
This provides a simple explanation of the reported in
\cite{cz} lower than maximal Einstein-Podolsky- Rosen (EPR)
correlations (and, accordingly, weak or no violations of Bell-type
inequalities).

\medskip

 This work is supported by a grant from the Technion Graduate
School. It is a pleasure to acknowledge discussions with Israel
Klich, Ady Mann, Asher Peres and Yoshua Zak. Comments and
criticism of the anonymous referee  significantly improved the
presentation.

\end{document}